\documentclass[3p,times]{elsarticle}

\usepackage{lineno,hyperref}
\usepackage{floatrow}

\modulolinenumbers[5]

\journal{Annals of Physics}

\biboptions{sort&compress}

\usepackage{amssymb,amsmath,textcomp}
\usepackage{float}
\usepackage{bm}

\usepackage{graphicx}
\usepackage{epsfig,epsf}
\usepackage{epstopdf}

\usepackage{makeidx}

\usepackage{color}
\usepackage{hyperref}
\usepackage{caption}
\usepackage{subcaption}

\usepackage{setspace}

\usepackage[most]{tcolorbox}
\usepackage{tikz,lipsum,lmodern} 

\definecolor{LY}{RGB}{255,255,200}
\definecolor{NG}{rgb}{0.9,0.96,0.8}
\colorlet{LG1}{green!15!yellow!30}

\makeindex

\newcommand{\beq}{\begin{eqnarray}}
	\newcommand{\eeq}{\end{eqnarray}}

\def\be{\begin{equation}}
	\def\ee{\end{equation}}

\def\tr{\mbox{Tr}}

\newcommand\eq[1]{Eq.~(\ref{#1})}

\numberwithin{equation}{section}

\begin{document}

	\begin{frontmatter}

		\title{\centering{Casimir and Helmholtz forces in one-dimensional Ising model with Dirichlet (free) boundary  conditions}} 
	
	\author[DMD,DMD2,JR]{D. M. Dantchev\corref{CorrespondingAuthor}}
	\ead{daniel@imbm.bas.bg}
	\author[NT]{N. S. Tonchev}
	\ead{nicholay.tonchev@gmail.com}
	\author[JR]{J. Rudnick}
	\ead{jrudnickucla@gmail.com}
	
	\address[DMD]{Institute of
		Mechanics, Bulgarian Academy of Sciences, Academic Georgy Bonchev St. building 4,
		1113 Sofia, Bulgaria}
	\address[DMD2]{Max-Planck-Institut f\"{u}r Intelligente Systeme, Heisenbergstrasse 3, D-70569 Stuttgart, Germany}
	\address[NT]{Institute of Solid State Physics, Bulgarian Academy of Sciences,1784 Sofia, Bulgaria} 
	\address[JR]{Department of Physics and Astronomy, University of California, Los Angeles, CA 90095}
	
	\cortext[CorrespondingAuthor]{Corresponding author}
	

		
		\begin{abstract}
			Attention in the literature has increasingly turned to the issue of the dependence on  ensemble and boundary conditions of fluctuation-induced forces. 
			 We have recently investigated this problem in the one-dimensional Ising model with periodic and antiperiodic boundary conditions (Annals of Physics {\bf 459}, 169533 (2023)). Significant variations of the behavior of Casimir and Helmholtz forces was observed, depending on both ensemble and boundary conditions.  Here we extend our study by considering  the problem in the important case of Dirichlet  (also termed free, or missing neighbors) boundary conditions. The advantage of the mathematical formulation of the problem in terms of Chebyshev polynomials is  demonstrated and, in this approach, expressions for the partition functions in the canonical and the grand canonical ensembles are presented. We prove analytically that the Casimir force is attractive for all values of temperature and external ordering field, while the Helmholtz force can be both attractive and repulsive.			 
		\end{abstract}
		
		\date{\today}
		
	\end{frontmatter}

\section{Introduction}	
We start by recalling the definitions of the fluctuation induced forces - the Casimir force in the grand canonical ensemble (GCE), and the analogous Helmholtz force in the canonical ensemble (CE). 

 We envisage a  $d$-dimensional system with a film geometry $\infty^{d-1}\times L$, $L\equiv L_\perp$, and with boundary conditions $\zeta$ imposed along the spatial direction of finite extent $L$.  Take ${\cal F}_{ {\rm tot}}^{(\zeta)}(L,T,h)$ to be the total free energy of such a system within the GCE, where $T$ is the temperature and $h$ is the magnetic field. Then, if   $f^{(\zeta)}(T,h,L)\equiv \lim_{A\to\infty}{\cal F}_{ {\rm tot}}^{(\zeta)}/A$  is the free energy per area $A$ of the system, one can define the Casimir force for critical systems in the grand-canonical $(T-h)$-ensemble, see, e.g. Ref. \cite{Krech1994,BDT2000,MD2018,DD2022,Gambassi2023}, as: 
\begin{equation}
	\label{CasDef}
	\beta F_{\rm Cas}^{(\zeta)}(L,T,h)\equiv- \frac{\partial}{\partial L}f_{\rm ex}^{(\zeta)}(L,T,h)
\end{equation}
where
\begin{equation}
	\label{excess_free_energy_definition}
	f_{\rm ex}^{(\zeta)}(L,T,h) \equiv f^{(\zeta)}(L,T,h)-L f_b(T,h)
\end{equation}
is the so-called excess (over the bulk) free energy per area and per $\beta^{-1}=k_B T$.

Along these lines, we define the corresponding  fluctuation induced   Helmholtz force \cite{DR2022,Dantchev2023b} in the canonical $(T-M)$-ensemble, where $M$ is the fixed value of the total magnetization:
\begin{equation}
	\label{HelmDef}
	\beta F_{\rm H}^{(\zeta)}(L,T,M)\equiv- \frac{\partial}{\partial L}f_{\rm ex}^{(\zeta)}(L,T,M)
\end{equation}
and
\begin{equation}
	\label{excess_free_energy_definition_M}
	f_{\rm ex}^{(\zeta)}(L,T,M) \equiv f^{(\zeta)}(L,T,M)-L f_H(T,m).
\end{equation}
Here, the average magnetization $m=\lim_{L, A\to \infty}M/(LA)$, and $f_H(T,m)$ is the Helmholtz free energy density of the ``bulk'' system. In the remainder of this article we will take $L=N a$, where $N$ is an integer number, and for simplicity we 
set $a=1$, i.e. all lengths will be measured in units of the lattice spacing $a$. 

We will show, via exact results for the Ising chain, that the  fluctuation-induced Helmholtz force under Dirichlet boundary conditions (DBC's), as in the recently considered case of periodic (PBC's) and antiperiodic (ABC's) boundary conditions \cite{Dantchev2023b},  has behavior very different from that of the Casimir force. Critical Casimir and Helmholtz forces are the thermodynamic analogues of the quantum electrodynamical Casimir force. In critical systems one studies the effects due to the massless excitations of the order parameter, while in the quantum case on studies the fluctuations of the electromagnetic field. Given the difference between GCE and CE 
for finite systems, it is reasonable to anticipate that Casimir and Helmholtz forces will have, in general,
different behavior for the same geometry and boundary conditions. On the basis of knowledge accumulated in the course of considering different boundary conditions it is normally assumed that if the boundary conditions are the same, or similar, on both confining the system surfaces between which they act, the  Casimir force is attractive---in the opposite case, repulsive. However, it seems that there is no such a rule for Helmholtz forces. Based on the results reported in \cite{DR2022} for PBC's and in \cite{Dantchev2023b} for ABC's one concludes that while the rule os obeyed for the Casimir force---attractive for PBC's and repulsive for ABC's---the Helmholtz force changes sign as a function of $T$ and $M$ for both PBC's and ABC's. The case of DBS's for the Casimir force with $h=0$ has been studied in \cite{RZSA2010}, in which it was shown  that the Casimir force trivially vanishes. Results for the Casimir force with $h\ne0$ and for the Helmholtz force with DBC have, as far as we are aware, not been reported in the available literature. It should be noted that the precise behavior of Helmholtz forces has not as yet been the subject of thorough and systematic study, and just a small number of results are as yet available. 

In many circumstances the
experimental setup of the CE is naturally realized in confined systems \cite{Henkel2008,WAP2017,DR2022}, which  justifies the need for a canonical statistical  representation of these systems. The vast majority of theoretical studies of the canonical partition function and the related Helmholtz free energy in  {\it finite -size systems} have been performed in the framework  of different  \textit{approximation} schemes: within the mean field approximation based on the Landau-Ginzburg free energy \cite{GVGD2016,GGD2017,RSVG2019}, via the $\varepsilon$ expansion technique \cite{ET87}, or using numerical methods such as Monte Carlo simulations of the three-dimensional Ising model; see Refs. \cite{ET87,BHT2000,shen2020finite} and reverences therein.
We note that in many cases  the
application  of the equilibrium one dimensional Ising model on the rigorous level  with $M$ fixed is required.  Examples are: binary allows or binary liquids, some one dimensional van-der-Waals materials see e.g. Ref. \cite{Balandin2022a} and refs. therein, Feynman checkers (known also as Hadamard walk) Ref. \cite{Skopenkov2020} and refs. therein, which demonstrate an existing interest in the matter. Further examples  are to be found in  \cite{antal2004probability,denisov2005domain,Nandhini2009,GarciaPelayo2009,kofinger2010single,taherkhani2011investigation,wang2019solving,Ferreira2023}. 
Though the finite-size consideration in conjunction with different boundary conditions entails significant mathematical efforts,  the most attractive feature of the model still remains: for many quantities of interest closed-form answers are available.  

The one-dimensional 
Ising model in the GCE is widely discussed in textbooks on Statistical 
Mechanics \cite{B82,H87,K2007,PB2011,BH2019}.  However, as a theoretical set up to elucidate  boundary and size effects the CE has been consistently, and undeservedly, ignored in the literature. 
It it fair to say that one of the reasons for this is that the case of GCA requires relatively simple mathematics while the CE is based on more involved mathematical techniques. Specifically, it turns out that the use of usual standard transfer matrix approach can also be exploited, but only in combination with 
the Chebishev polynomials \cite{Dantchev2023b} and generalized Gauss  hypergeometric functions  \cite{DR2022,Dantchev2023b}.

In the present study, we will present exact results for the Casimir force in GCE and for the Helmholtz force in CE in the case in which the Ising chain is subjected to Dirichlet boundary conditions. This will serve to emphasize the role of the boundary conditions for finite-size scaling behavior of the system. We will discuss the relation between the CE and GCE at finite size and the corresponding finite size corrections to the bulk quantities.

The article is organized as follows. Sec. \ref{sec:free-energy-Casimir-forec} contains the derivation of the Gibbs free energy and the behavior of Casimir force within the grand canonical ensemble with DBC's. The corresponding derivation for the Helmholtz free energy and the behavior of Helmholtz force within the canonical ensemble with DBC's are reported in Sec. \ref{sec:Helmholtz-force-and-energy}. Section  \ref{sec:transfer-matrix-method} is devoted on  the derivation of $Z^{(D)}_C(N,K,M)$ via the transfer matrix method, elucidates few relations between the Chebyshev polynomials and the Gauss hypergeometric function.  The article ends with a section \ref{sec:conclusion} that summarizes the  results obtained and discusses their potential applicability. Most technical details needed for the derivations and the justifications of our statements are organized in series of appendices at the end of the article.

 \section{Gibbs free energy and Casimir force within the grand canonical ensemble with DBC's}
 \label{sec:free-energy-Casimir-forec}
 
 We will consider a one-dimensional Ising chain of $N$ spins $(S_i \pm 1, i=1,.., N)$
 with ferromagnetic coupling $J>0$ and its properties under various boundary conditions (BC's). The Hamiltonian of the model is given by
 \begin{equation}
 	H=-J\sum_{i=1}^{N-1} S_i S_{i+1} - J_{BC}(S_{1}S_N) + h\sum_{i=1}^N S_i,
 \end{equation}
 where $J_{BC's}=J, -J, 0$ for periodic (PBC's), antiperiodic (ABC's) and Dirichlet-Dirichlet (DBC's) (also termed free, or missing neighbors), boundary conditions, respectively.

The
partition function of the model with DBC's in the grand canonical ensemble is well known. It is usually derived
by the method of transfer matrix (see Eq.(2.14) in \cite{MW73}) which yields
\begin{eqnarray}
\label{1Di}
Z^{(D)}_{\rm GC}(N,K,h)=\lambda^{N-1}_{1}[\cosh(h)
+ A(h,K)] + \lambda^{N-1}_{2}[\cosh(h)
- A(h,K)],
\end{eqnarray}
 or, equivalently
\begin{eqnarray}
	\label{cu}
	Z^{(D)}_{\rm GC}(N,K,h)
	=\cosh(h)\bigg[\lambda^{N-1}_{1}(K,h)+\lambda^{N-1}_{2}(K,h)\bigg]+ A(K,h)\bigg[\lambda^{N-1}_{1}(K,h)-\lambda^{N-1}_{2}(K,h)\bigg].
\end{eqnarray}
Here $\lambda_{1,2}$ are the two real eigenvalues of the corresponding transfer matrix {\bf T}
\begin{equation} 
	\label{l120}
	\lambda_{1,2}(K,h)=
	e^K\cosh(h) \pm \sqrt{e^{-2K}+
		\left[e^K\sinh(h)\right]^2 }\equiv \sqrt{2\sinh(2K)}\,\bigg[z\pm \sqrt{z^2-1}\bigg],
\end{equation}
 and 
\begin{equation}
	\label{eq:A-definition}
 A(K,h)=\frac{\sinh^2(h)+e^{-2K}}{\sqrt{\sinh^2(h)+e^{-4K}}}
=\frac{e^K\sinh^2(h)+e^{-K}}{\sqrt{2\sinh(2K)}\sqrt{z(K,h)^2 -1}}.
\end{equation}
We have also introduced the auxiliary variable $z(k,h)$:  
\begin{equation}
	\label{Nsv}
	z=  z(K,h)=\frac{e^K\cosh(h)}{\sqrt{2\sinh(2K)}}\; \equiv\;\frac{\tr{({\bf T})}}{2\sqrt{\det{({\bf T})}}}>1,\quad\forall h\in [0,\infty), K>0,
\end{equation}
the  importance of which will become clear later.  
In the above expressions  $K\equiv\beta J$ is the dimensionless coupling, and $\beta=1/(k_B T)$ is the inverse temperature. Since $\lambda_1(K, h)=\lambda_1(K, -h)$ and $\lambda_2(K, h)=\lambda_2(K, -h)$, one has $Z^{\rm (D)}(N,K,h)=Z^{\rm (D)}_{\rm GC}(N,K,-h)$. Because of this, we will henceforth assume without loss of generality that $h\ge 0$. 

In order to establish the connection between these expressions and the partition function in the canonical ensemble and to explore the relationship between them, it is useful to present these results in terms of Chebyshev polynomials \cite{Mason2002,rivlin2020}).

We start with the identity 
\begin{eqnarray}
\label{curb}
\lambda_1^{N-1}(K, h) \pm \lambda_2^{N-1}(K, h) &=&
\left(\sqrt{2\sinh(2K)}\right)^{N-1}\nonumber\\
\times \bigg\{
\left[z(K,h) + \sqrt{\left[z(K,h)\right]^2-1}\right]^{N-1} &\pm&
\left[z(K,h) - \sqrt{\left[z(K,h)\right]^2-1}\right]^{N-1}
\bigg\}.
\end{eqnarray}
By virtue of the definition of the Chebyshev polynomials of  first kind $T_{N}(z)$:
\begin{equation}
\label{cp11}
\left[(z+\sqrt{z^2-1})^{N-1} + (z-\sqrt{z^2-1})^{N-1}\right]=2T_{N-1}(z), \quad z\in {\mathbb C}.
\end{equation}
 the expression in the curly  brackets, Eq.\eqref{curb}, with sign ``+" is exactly $T_{N-1}(z)$ and thus
\begin{equation}
\label{eq:free_energy_pern1}
\lambda^{N-1}_{1}(K,h) +
 \lambda^{N-1}_{2}(K,h)
=2\left(\sqrt{2\sinh(2K)}\right)^{N-1} T_{N-1}\left(z(K,h\right)). \quad 
\end{equation}
The term with  sign ``-" in the curly  brackets, Eq.\eqref{curb}, may be represented in terms of  the Chebyshov polinomials of the second kind as a consequence the identity  
\begin{equation}
\label{cp116}
\left[(z+\sqrt{z^2-1})^{N-1} -(z-\sqrt{z^2-1})^{N-1}\right]=2\sqrt{z^2-1}U_{N-2}(z),\quad z\in {\mathbb C}
\end{equation}
which allows us to obtain 
\begin{equation}
\label{eq:free_eney_pern1}
 \lambda^{N-1}_{1}(K,h) -
\lambda^{N-1}_{2}(K,h)
=2\left(\sqrt{2\sinh(2K)}\right)^{N-1}\sqrt{z(K,h)^2-1}U_{N-2}(z(K,h)).
\end{equation}
Thus, for the grand canonical partition function with DBC's we find:
\begin{eqnarray}
		\label{GKD}
 Z^{(D)}_{\rm GC}(N,K,h)
&=&2\left[\sqrt{2\sinh(2K)}
 \right]^{N-1}\\
 && \times \bigg\{\cosh(h) T_{N-1}\bigg(z(K,h\bigg)
+\,\frac{e^K\sinh^2(h)+e^{-K}}{\sqrt{2\sinh(2K)}} U_{N-2}
		\bigg(z(K,h)\bigg)\bigg\}. \nonumber
\end{eqnarray}

Connections between $Z^{(D)}_{\rm GC}(N,K,h)$ and the partition functions of the systems with periodic and antiperiodic boundary conditions are presented in \ref{sec:GC-DD-relations}.

\subsection{On the critical behavior of the infinite Ising chain and the leading finite-size corrections}

The infinite one-dimensional Ising chain with short-ranged interactions has at $T=0$ a critical point with essential singularities. The free energy density of the this chain  is 
\begin{equation}
	\label{eq:bulk-gibbs-free-energy}
	\beta f_b(K,h)=-\ln \lambda_{1}(K,h),
\end{equation}
Knowing the behavior of the correlation length, see, e.g.,  Ref. \cite[p. 36, Eq. 2.2.15]{B82} 
\begin{equation}
	\label{eq:scaling_length_Ising}
	\xi^{-1}(K,h)=\ln \left[\lambda_1(K,h)/\lambda_2(K,h)\right],
\end{equation}
one can easily identify the scaling variables.  First, it is clear that $\xi(K,h)$ diverges when $\lambda_2(K,h)\to\lambda_{1}(K,h)$. Obviously, this happens when $h\to 0$ and $K\to\infty$.  Defining
\be
\label{eq:xit_Ising}
\xi_t\equiv\xi(K,0)\simeq \frac{1}{2}e^{2K}, \;\mbox{when}\; K\gg 1, \mbox{and}\; \xi_h\equiv\lim_{K\to \infty}\xi(K,h)\simeq \frac{1}{2h}, \; \mbox{when} \; h\ll 1,
\ee
for the scaling variables one identifies
\be\label{eq:scaling_variables_Ising}
x_t=N/\xi_t= 2 N e^{-2K}, \quad \mbox{and} \quad x_h =N/\xi_h=2 N h. 
\ee
Thus, in terms of these scaling variables the correlation length, the bulk magnetization, and the bulk Gibbs free energy density in the limit $K\gg 1$ are 
\be
\label{eq:corr_length_scaling}
\xi(K,h)=\frac{N}{\sqrt{x_h^2+x_t^2}}, \quad m_b(K,h)=\frac{x_h}{\sqrt{x_h^2+x_t^2}}, \quad \beta f_b(K,h)= -K-\frac{1}{4 N}\sqrt{x_h^2+x_t^2}.
\ee

In terms of $t=\exp(-2K)$ and $h$, one can define the usual scaling relations with 
\be
\label{eq:crit_exponents}
\alpha=\gamma=\nu=\eta=1,\; \beta=0,\;\delta=\infty, \; \mbox{but so that}\; \beta \delta=1. 
\ee

The free energy density of the finite chain is
\begin{equation}
	\label{eq:chain-gibbs-free-energy}
	\beta f^{(D)}_{\rm GC}(N,K,h)=-\frac{1}{N} \ln Z^{(D)}_{\rm GC}(N,K,h).
\end{equation}
Then, from \eq{1Di} we immediately obtain 
\begin{equation}
	\label{eq:free-energy-finite-Dirichlet-prepared-for-corrections}
	\beta f^{(D)}_{\rm GC}(N,K,h) = \beta f_b(K,h) +\frac{1}{N}\beta f_{\rm surface}(K,h)+ \beta \Delta f_N(K,h),
\end{equation}
where the "surface" free energy is
\begin{equation}
	\label{eq:surface-free-energy}
	\beta f_{\rm surface}(K,h)=\ln \lambda_1(K,h)-\ln[A(K,h)+\cosh (h)],
\end{equation} 
while the remnant part $\beta \Delta f_N(K,h)$ is 
\begin{equation}
	\label{eq:remnant-part}
	\beta \Delta f_N(K,h)=-\frac{1}{N}\ln\left\{1+\exp \left[-\frac{N-1}{\xi (K,h)}\right] \;\frac{\cosh (h)-A(K,h) }{\cosh (h)+ A(K,h)}\right\}.
\end{equation}
\eq{eq:chain-gibbs-free-energy} is in a full agreement with Ref. \cite[Eq. (2.27)]{MW73}. Note that $\beta \Delta f_N(K,h)$ is exponentially small for $N\gg 1$ when $\xi (K,h)={\cal O}(1)$. The only  exception is the case when $N\propto \xi (K,h)$, i.e., $N/\xi (K,h)={\cal O}(1)$. The last relation defines the so-called finite-size scaling region, described by scaling variables as given in \eq{eq:scaling_variables_Ising}. 

\subsection{On the behavior of the Casimir force}

For the behavior of the Casimir force, let us start with the case $h=0$. It is easy to check that then $A(K,h=0)=1$ and only $\lambda_{1}(K,h)$ gives a contribution to the statistical sum. This leads to  excess free energy that is independent of $N$ and, therefore, as reported in Ref. \cite{RZSA2010}
\begin{equation}
	\label{eq:Casimir-at-zero-field}
	\beta F_{\rm Cas}^{(D)}(N,K,h=0)=0.
\end{equation}
The situation changes if $h\ne 0$. Then $F_{\rm Cas}^{(D)}(N,K,h\ne 0)\ne 0$ and, from Eqs. \eqref{CasDef}, 	\eqref{excess_free_energy_definition} and  \eq{eq:chain-gibbs-free-energy}, one derives
\begin{equation}
	\label{eq:Casimir-nonzero-field}
	\beta F_{\rm Cas}^{(D)}(N,K,h)=-\frac{1}{N} \frac{N}{ \xi (K,h)} \frac{1}{r(K,h) \exp \left[(N-1)/\xi (K,h)\right]+1}, \quad \mbox{where} \quad r[K,h]:=\frac{\cosh (h)+A(K,h)}{\cosh (h)-A(K,h)}.
\end{equation}
For $K>0$ and $h\ne 0$, since  $0<A(K,h)<\cosh(h)$ (for an elementary proof, see \eqref{Ineq}), one has $r[K,h]>0$. Thus $\beta F_{\rm Cas}^{(D)}(N,K,h)<0$ for any $K>0$ and $h\ne0$. 

Let us now consider the behavior of the force in the scaling regime. By definition, the scaling function of the Casimir force $X_{\rm Cas}(x_t,x_h)$ is 
\begin{equation}
	\label{eq:scaling-function-Casimir-definition}
	\beta F_{\rm Cas}^{(D)}(N,K,h)=\frac{1}{N} X_{\rm Cas}(x_t,x_h).
\end{equation}
Explicitly, from \eq{eq:scaling_variables_Ising} and \eq{eq:Casimir-nonzero-field} one has 
\begin{equation}
	\label{eq:scaling-function-Casimir}
	X_{\rm Cas}(x_t,x_h)=-\frac{\sqrt{x_h^2+x_t^2}}{r(x_t,x_h) \exp \left(\sqrt{x_h^2+x_t^2}\right) +1},  \quad \mbox{where} \quad r(x_t,x_h)= \frac{\sqrt{x_h^2+x_t^2}+x_t}{\sqrt{x_h^2+x_t^2}-x_t}.
\end{equation}
Obviously, $X_{\rm Cas}(x_t,x_h)<0$. Furthermore,  $X_{\rm Cas}(x_t,x_h)$ decays exponentially when $x_h^2+x_t^2\gg 1$.

\begin{figure}[h!]
	\centering
	\includegraphics[width=3.0in]{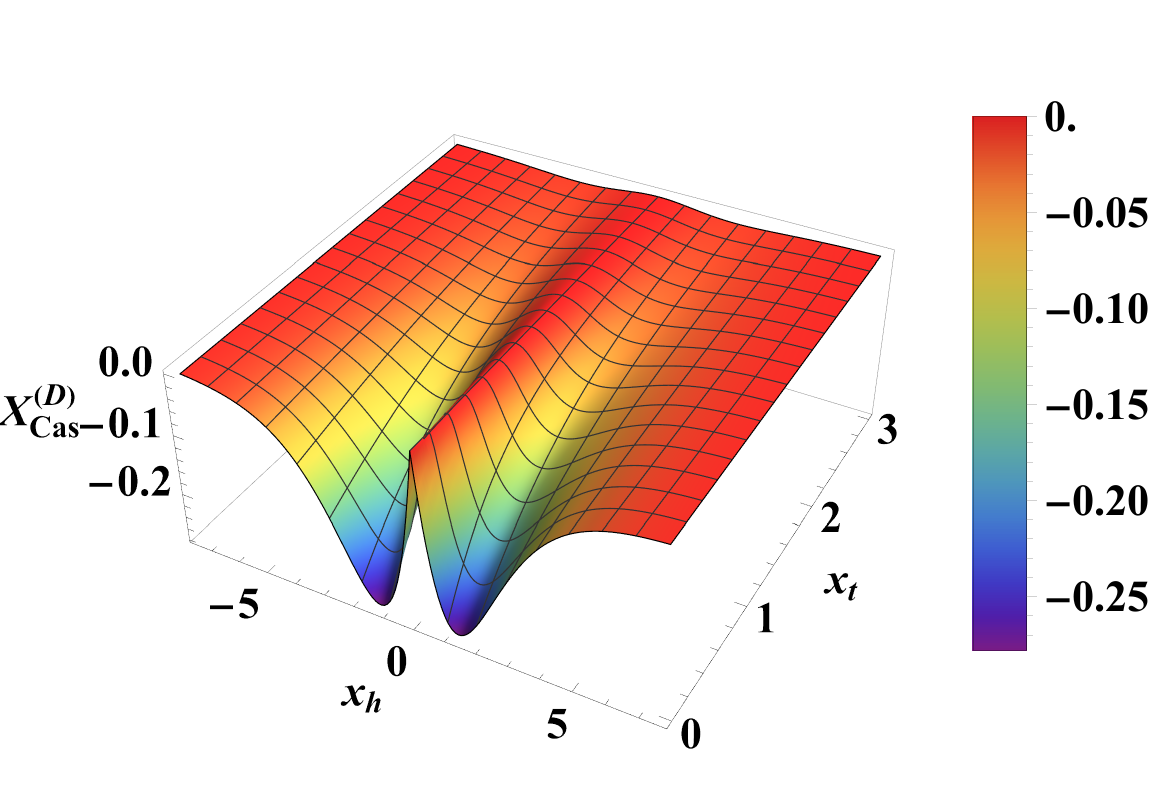} \quad
	\includegraphics[width=3.0in]{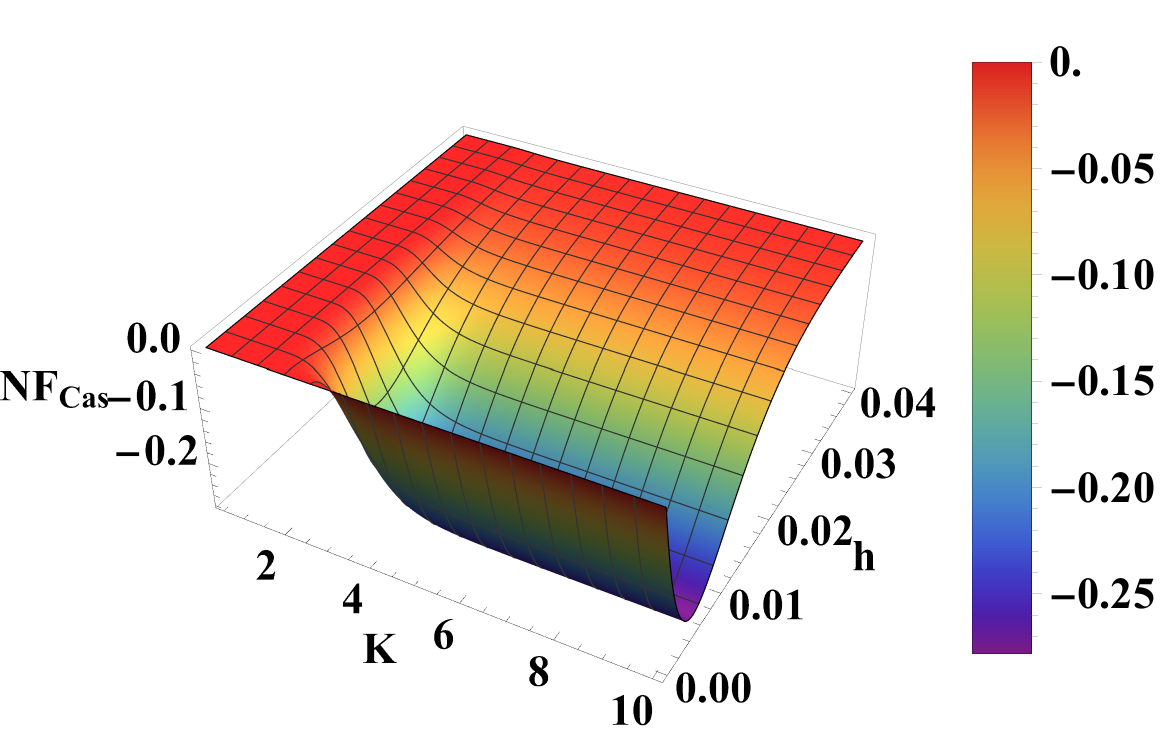}
	\caption{On the left panel: The behavior of the  scaling function $X_{\rm Cas}^{(D)}(x_t,x_h)$  of the Casimir force as a function of the scaling variables $x_t$ and $x_h$. We observe that the function is negative  for all values of $x_t$ and $x_h$. On the right panel: The behavior of the function $N F_{\rm Cas}^{(D)}$ with $N=100$.}
	\label{fig:3D-Casimir}
\end{figure}

The behavior of the function $N F_{\rm Cas}^{(D)}$ with $N=100$ and the scaling function $X_{\rm Cas}^{(D)}(x_t,x_h)$  of the Casimir force are visualized in Fig. \ref{fig:3D-Casimir}. We observe that the force is always \textit{attractive}. It is  symmetric with respect to the sign of $x_h $, as  must be the case.

\section{Helmholtz free energy and Helmholtz force within the canonical ensemble with Dirichlet BC's}
\label{sec:Helmholtz-force-and-energy}

In a recent letter \cite{DR2022} for an Ising chain of $N$ spins (particles) and total magnetization $M$, with the aid of a  combinatorial analysis  an exact expression was derived for the partition function in the canonical (i.e., fixed $M$) ensemble. The result is expressed in closed form in terms of the generalized Gauss hypergeometric functions:
	\begin{eqnarray}
		\label{DRF}
	Z^{(D)}(N,K,M) 
	& = & e^{K (N-1)} \Bigg[2 e^{-2 K} \, _2F_1\left(\frac{1}{2} (-M-N+2),\frac{1}{2}
	(M-N+2);1;e^{-4 K}\right) \nonumber \\ &&-\frac{1}{2} e^{-4 K} (M-N+2) \, _2F_1\left(\frac{1}{2}
	(-M-N+2),\frac{1}{2} (M-N+4);2;e^{-4 K}\right) \nonumber \\ && + \frac{1}{2} e^{-4 K} (M+N-2) \,
	_2F_1\left(\frac{1}{2} (-M-N+4),\frac{1}{2} (M-N+2);2;e^{-4 K}\right)\Bigg],   \label{eq:DDbc-main-text}
\end{eqnarray}
	where 
\begin{equation}
	\label{dhf}
	_2F_1\left(\alpha,\beta;\gamma;z\right)=
	\sum_{k=0}^{\infty}\frac{(\alpha)_k(\beta)_k}
	{(\gamma)_k}\frac{z^k}{k!}
\end{equation}
is the generalized Gauss hypergeometric function with parameters  $\alpha, \beta, \gamma $, and  $\gamma \neq  0, -1, -2, ... $. 
We recall that if in \eq{dhf} $\alpha$ and/or $\beta$ are
	negative integers, which is the case of the present study, 
	the Gauss series reduces to a hypergeometric polynomial, so that convergence of the series is not an issue.

Below, using this expressions, we will derive the fluctuation induced force pertinent to this ensemble, for which in \cite{DR2022} the term Helmholtz force has been coined. In \ref{rpf} it is shown that the representation of $	Z^{(D)}(N,K,M)$  given by Eq.\eqref{DRF} can be simplified to 
\begin{align}
	\label{sl2j}
	Z^{(D)}_C(N,K,M)&= e^{K(N-4)}\bigg[4\sinh(K)\,_2F_1\bigg(\frac{1}{2}(M-N+2),\frac{1}{2}(-M-N+2);1;e^{-4K}\bigg)
	+\nonumber\\
&	Ne^{-K}\,_2F_1\bigg(\frac{1}{2}(M-N+2),\frac{1}{2}(-M-N+2);2;e^{-4K}\bigg)\bigg].
\end{align}
It is worth noting the striking similarity between equations Eq.\eqref{sl2j} for DBC's and
\begin{align}
	\label{sl2jh}
	Z^{\rm (anti)}_C(N,K,M)&= e^{K(N-4)}\bigg[4\sinh(2K)\;_2F_1\bigg(\frac{1}{2}(M-N+2),\frac{1}{2}(-M-N+2);1;e^{-4K}\bigg)
	+\nonumber\\
&	Ne^{-2K}\;_2F_1\bigg(\frac{1}{2}(M-N+2),\frac{1}{2}(-M-N+2);2;e^{-4K}\bigg)\bigg],
\end{align}
for ABC's in the case of the canonical ensemble and the distinction of the corresponding partition functions in the grand canonical ensemble. 

In order to determine the behavior of the Helmholtz force we need to know the Helmholtz free energies in the finite and infinite Ising chains with a given average magnetization $m$. 

The bulk Helmholtz free energy has been reported in Ref. \cite{Dantchev2023b}. It is obtained with the use of a Legendre transformation from the bulk Gibbs free energy (the bulk ensembles are equivalent) and reads 
\begin{align}
	\label{eq:Helmholtz_free_energy_bulk_explicit}
	\beta a_b(K,m) = K + \frac{1}{2}\ln(1 - m^2) - \ln\left[1 + \sqrt{m^2 + e^{4 K} (1 - m^2)}\right]  + \; m \sinh ^{-1}\left(\frac{e^{-2 K} m}{\sqrt{1-m^2}}\right).
\end{align}
For the scaling behavior of $\beta a_b(K,m)$, i.e., for $K\gg 1$, from 	\eq{eq:Helmholtz_free_energy_bulk_explicit} one obtains 
\begin{equation}
	\label{eq:ab-scaling}
	\beta a_b(K,m)\simeq -K -\exp[-2K]\sqrt{1-m^2}=-K-\frac{1}{2N}x_t\sqrt{1-m^2}, \quad K \gg 1. 
\end{equation}

The Helmholtz free energy density of the Ising chain with fixed $M$ is by definition 
\begin{equation}
	\label{eq:Helmholtz-free-energy-definition}
	\beta a(N,K,M)=-\frac{1}{N} \ln Z^{(D)}_C(N,K,M). 
\end{equation}
Let us first derive the behavior of the partition function $Z^{(D)}_C(N,K,M)$ in the scaling regime $x_t={\cal O}(1)$ with $M=Nm$. 
To proceed, we use the  the asymptotic expansion obtained in Ref. \cite[Eq. D6]{Dantchev2023b} (there $-a=\frac{1}{2}(Nm-N+2),\, -b=\frac{1}{2}(-Nm-N+2))$:
\begin{align}
	\label{eq:expansion-D6}	
	& _2F_1(\frac{1}{2}(Nm-N+2),\frac{1}{2}(-Nm-N+2);\gamma;e^{-4K}) \\
	&\simeq (\gamma-1)! \left(\frac{1}{4} \sqrt{1-m^2}\, x_t\right)^{1-\gamma}
	\left[I_{\gamma-1}\left( \frac{1}{2}\sqrt{1-m^2}\, x_t\right)
	+O(N^{-1})\right]. \nonumber
\end{align}
Here $\gamma={\cal }O(1)$, $\exp[-4K]\ll 1$, and $I_{\nu}(z)$ is the modified Bessel function of first kind: 
\begin{equation}
	\label{DefB}
	I_{\nu}(z)=\sum_{k=0}^{\infty}
	\frac{1}{k! \Gamma(\nu+k+1)}\left(z/2\right)^{\nu+2k} \simeq\left\{\begin{array}{cc}\left(z/2\right)^{\nu}/\Gamma(\nu+1), & z\to 0,\\ 
		&\\e^z/\sqrt{2\pi z}, & z\to \infty. \end{array}\right.
\end{equation}
Inserting the asymptotic expansion of Eq. \eqref{eq:expansion-D6}, with $\gamma=1$ and $\gamma=2$   into \eq{sl2j}   and summing the results we arrive at 
\begin{equation}
	\label{slb2}
	Z^{(D)}_C(N,K,m)= \frac{1}{N}
	e^{K(N-1)} x_t \bigg[I_0\left(\frac{1}{2}
	\sqrt{1-m^2}\, x_t\right)+
	\frac{ I_1\left(\frac{1}{2} \sqrt{1-m^2} x_t\right)}{\sqrt{1-m^2}}\bigg]\left(1+{\cal O}(N^{-1})\right).
\end{equation}
The conditions $x_t\to 0$ and $x_t\to \infty$ imply the asymptotes
\begin{align}
	\label{Dscw}
	Z^{(D)}_C(N,K,m)	\approx \frac{1}{N} e^{K(N-1})x_t\, \left\{
	\begin{array}{cc} \bigg[2+ \left(1-m^2\right) x_t^2/32\,\bigg], &
		 x_t\to 0,\\ 
		&  \\
		\bigg[\frac{\exp{\left(\sqrt{1-m^2} x_t/2\right)} }{\sqrt{\pi } \sqrt{1-m^2} x_t}\bigg]\,\bigg[1
		+ \left(\sqrt{1-m^2} x_t\right)^{-1/2}\bigg],
		& x_t\to \infty. \end{array}
		\right. 
\end{align}

\begin{figure}[h!]
	\centering
	\includegraphics[width=3.0in]{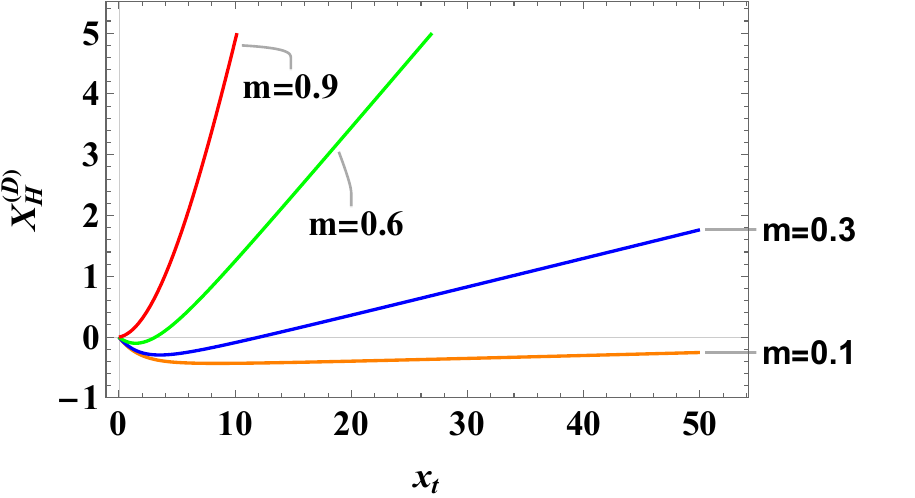} \quad
	\includegraphics[width=3.0in]{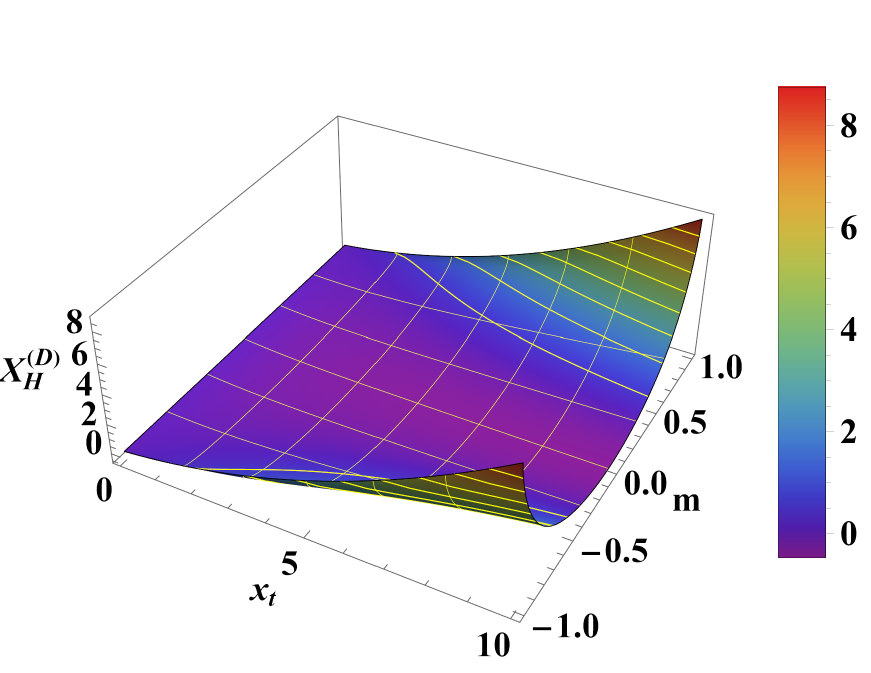}
	\caption{On the left panel: The behavior of the  scaling function $X_{\rm H}^{(D)}(x_t,m)$  of the Helmholtz force as a function of the scaling variable $x_t$ and $m$ for $m=0.1, m=0.3, m=0.6$ and $m=0.9$. We observe that the function is negative  for small values of $x_t$ for $m^2<3/4$ (see the $x_t\to 0$ asymptote in \eq{eq:asymptotes-of_helmholtz-scaling-function}). For $m^2>3/4$ and large values of $x_t$ the force is always repulsive. On the right panel: The behavior of the function $X_{\rm H}^{(D)}(x_t,m)$. We observe that, depending on the values of $x_t$ and $m$ the force can be both attractive \textit{and} repulsive. This behavior is remarkably different from that of the Casimir force which is always attractive for the same boundary conditions. }
	\label{fig:3D-Helmholtz}
\end{figure}

\begin{figure}[h!]
	\centering
	\includegraphics[width=3.0in]{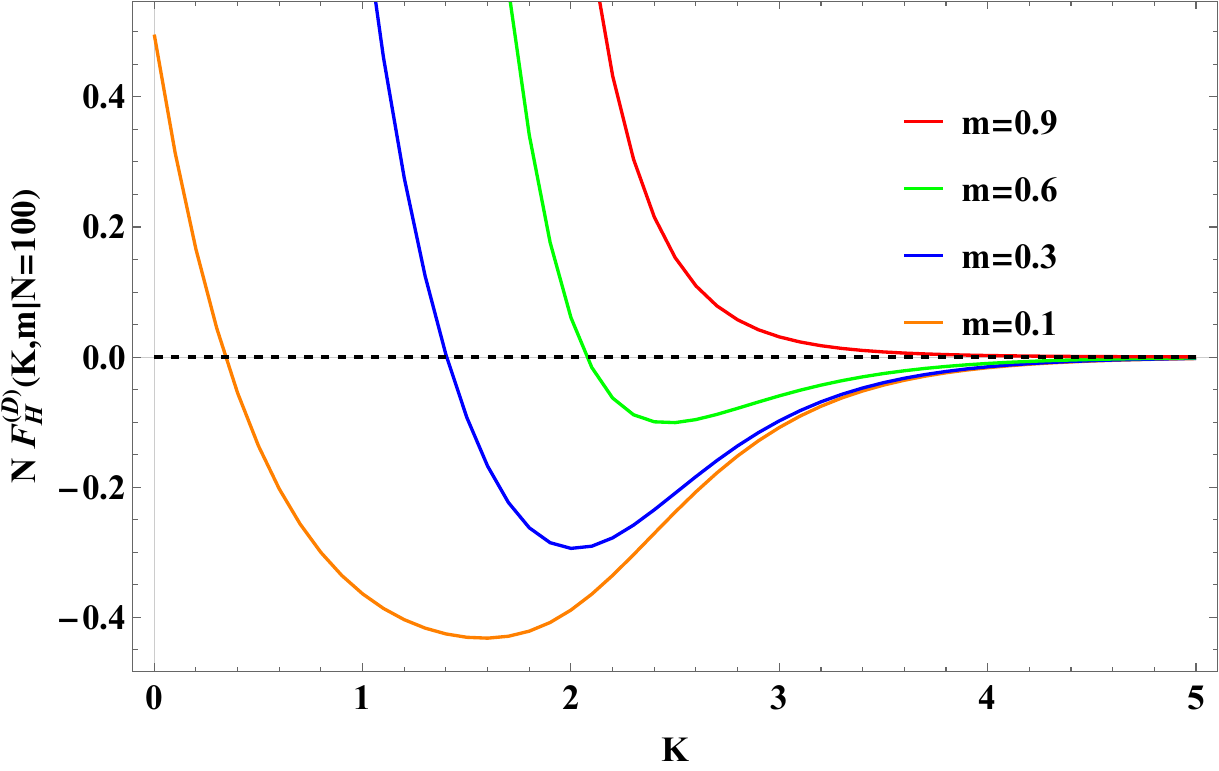}\quad
	\includegraphics[width=3.0in]{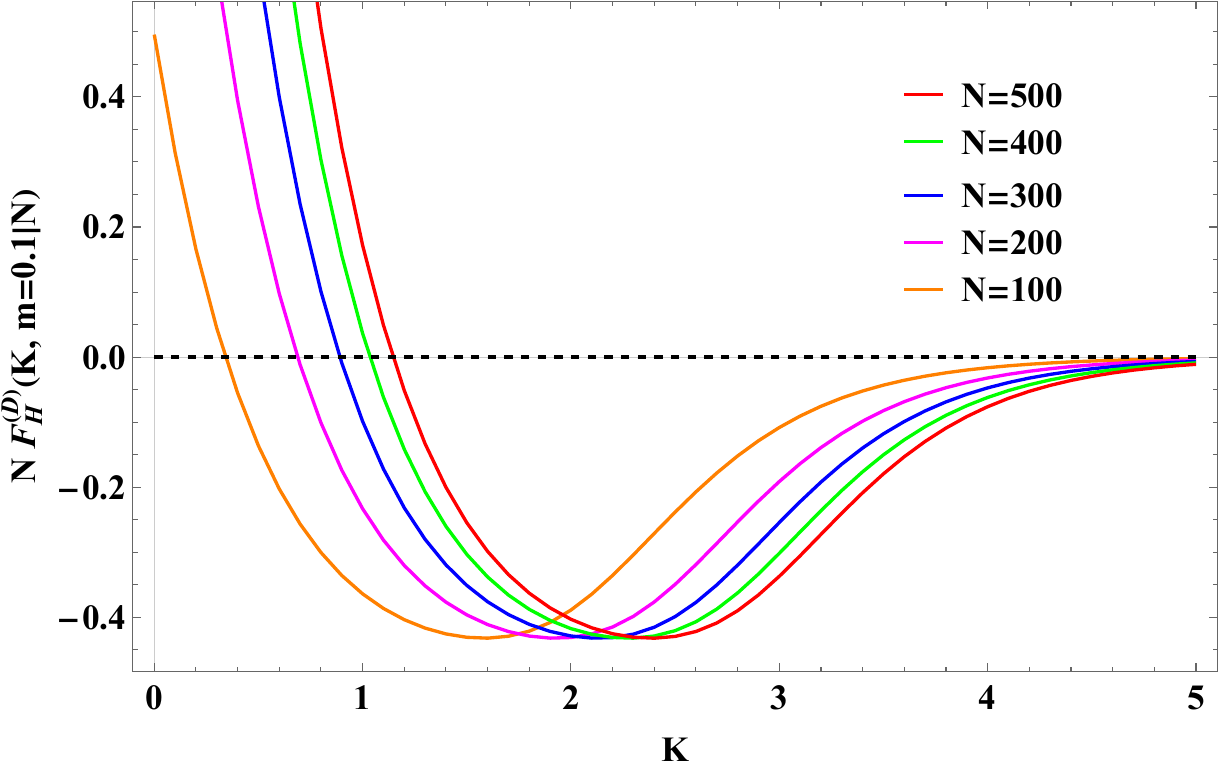}
	\caption{Left panel: The full $K$ behavior of the Helmholtz force illustrated fo $N=100$ and few values of $m$: $m=0.1, m=0.3, m=0.6$ and $m=0.9$. Right panel:  The $K$ dependence of the Helmholtz force for different values of $N$ for $N=100, 200, 300, 400$ and $N=500$. As we see - larger the $N$, stronger is the repulsion for small values of $K$.}
	\label{fig:full-K-Helmholtz}
\end{figure}

For the scaling function $X_H^{(D)}(x_t,m)$ of the Helmholtz force 
\begin{equation}
	\label{eq:Helmholtz-force}
	\beta F_{H}^{(D)}(N,x_t,m)=\frac{1}{N}X_H^{(D)}(x_t,m)
\end{equation}
from Eqs. \eqref{HelmDef}, \eqref{excess_free_energy_definition_M}, \eqref{eq:Helmholtz-free-energy-definition} and \eq{slb2}, we derive
\begin{align}
	\label{eq:scaling-function-of-Helmholtz-force}
	& X_H^{(D)}(x_t,m)= \\
	&	\frac{\left[\left(1-m^2\right) \left(1-\sqrt{1-m^2}\right) x_t-2 \left(m^2+1\right)\right] I_1\left(\frac{1}{2} \sqrt{1-m^2} x_t\right)+x_t\left(1-\left(1-m^2\right)^{3/2}\right) \sqrt{1-m^2}  I_0\left(\frac{1}{2} \sqrt{1-m^2} x_t\right)}{2 \left(1-m^2\right) \left[I_1\left(\frac{1}{2} \sqrt{1-m^2} x_t\right)+\sqrt{1-m^2} I_0\left(\frac{1}{2} \sqrt{1-m^2} x_t\right)\right]}. \nonumber
\end{align}
For its asymptotes we obtain
\begin{align}
	\label{eq:asymptotes-of_helmholtz-scaling-function}
	 X_H^{(D)}(x_t,m)\simeq \left\{\begin{array}{cc}
	\frac{1}{4} \left(1-2 \sqrt{1-m^2}\right) x_t+x_t^2/16+{\cal O}(x_t^{3}), & x_t\to 0\\
\frac{1}{2}\frac{m^2 }{\sqrt{1-m^2}} x_t-\frac{3/2-\sqrt{1-m^2}}{1-m^2}+{\cal O}(x_t^{-1}),	&  x_t \gg 1.
	\end{array} \right.
\end{align} 
The behavior of the scaling function of the Helmholtz force is visualized in Fig. \ref{fig:3D-Helmholtz}. The full $K$ behavior of the Helmholtz force is illustrated for $N=100$ and few values of $m$: $m=0.1, m=0.3, m=0.6$ and $m=0.9$ is shown in Fig. \ref{fig:full-K-Helmholtz}. We see  that for small $K$ the force is always repulsive. For moderate values of $K$ and $m$ the force changes sign from repulsive to attractive and then, for very large values of $K$ (corresponding to $x_t\to 0$) it levels off at $\beta F_H^{(D)}=0$.

\section{On the derivation of $Z^{(D)}_C(N,K,M)$ via transfer matrix method}	
\label{sec:transfer-matrix-method}

Here we demonstrate how to derive the combinatorial result \eq{eq:DDbc-main-text} or, equivalently,  \eq{sl2j} for the partition function in the canonical ensemble  via the transfer matrix method which, historically, was the one customarily used  for derivation of the properties of the Ising chain. It turns out that this is a non-trivial mathematical problem based on the algebra of the Gauss hypergeometric function and Chebyshev polynomials, that is interesting in itself.

We start with the  integral presentation of the Kronecker delta-function  
\begin{equation}
	\delta[S,M]=
	\frac{1}{2\pi}\int_{-\pi}^{+\pi}e^{i(S-M)\phi}d\phi,\quad S=\sum_{i=1} ^N S_{i},\quad S,M \in \mathbb{Z}.
\end{equation}
Then the canonical partition function  is given by
\begin{equation}
	Z^{(D)}_C(N,K,M)=\sum_{\{S_i\}_{\rm (D)}}
	e^{-\beta {\cal H}} \delta[S,M],\quad  \mbox{where} \quad {\cal H}=-J\sum_{i=1}^{N-1}S_iS_{i+1}.
\end{equation}
Here the symbol  $\{S_i\}_{\rm (D)}$ means that the set of spins obeys DBC's.
Further we have
\begin{eqnarray}
	\label{eq:transfer-matrix-dirichlet}
	Z^{(\rm D)}_C(N,K,M) &=&\frac{1}{2\pi}\int_{-\pi}^{\pi}{e^{-i M \phi}\left[\sum_{\{S_i\}^{\rm (D)}} e^{-\beta H+i \phi \sum_{i} S_i}\right]} d\phi=\frac{1}{2\pi}\int_{-\pi}^{\pi}{e^{-i M \phi}} \;Z^{(\rm D)}_{\rm GC}(N,K,i \phi)\,d\phi .\nonumber\\
\end{eqnarray}
Using \eqref{GKD}, one has 
\begin{eqnarray}
		\label{Zfree}
		Z^{(D)}_{GC}(N,K,i\phi)
		=2\left[\sqrt{2\sinh(2K)}
		\right]^{N-1}\bigg\{\cos(\phi) T_{N-1}\bigg(\tilde{z}(K,\phi)\bigg)+
	\,\frac{e^{-2K}-\sin^2(\phi)}{\cos(\phi) }\tilde{z}(K,\phi)U_{N-2}
		\bigg(\tilde{z}(K,\phi)\bigg)\bigg\}\nonumber\\
\end{eqnarray}
where 
\begin{equation}
	\label{Nsvo}
\tilde{z}(K,\phi)\equiv z(K,i\phi)=	 \frac{e^K\cos(\phi)}{\sqrt{2\sinh(2K)}},\quad K>0,\quad \phi \in [-\pi, \pi].
\end{equation}
Since the integrand in \eq{eq:transfer-matrix-dirichlet} is an even function of $\phi$---see \eq{Zfree}---we obtain
\begin{eqnarray}
	\label {Zd}
	&& Z^{(D)}_C(N,K,M)
	=\left[\sqrt{2\sinh(2K)}
	\right]^{N-1}\bigg\{\frac{4}{\pi}
	\int_{0}^{\pi/2}\cos(\phi)\cos( M \phi) T_{N-1}\bigg(\tilde{z}(K,\phi\bigg)d\phi+\nonumber\\
		&&+
		\frac{4}{\pi}\int_{0}^{\pi/2}\bigg[\frac{e^{-2K}-\sin^2(\phi)}{\cos(\phi) }
		\bigg]
		\cos ( M \phi)\tilde{z}(K,\phi)U_{N-2}
		\bigg(\tilde{z}(K,\phi)\bigg)d\phi\bigg\}.
	\end{eqnarray}	
Starting from the above expression and using  the recursion relations for the Chebyshev's polynomials \cite[8.941.1-8.941.3]{GR} one can show, see \ref{AF}, that 
\begin{equation}
	\label{eq:ZD-in-CE-integral}
Z^{(D)}_C(N,K,M)=	\frac{\sqrt{2 \sinh (2 K)}^N }{2 \cosh (K)}\left[\exp (-2 K)\, {\rm D}(N,M;e^{-4K})+{\rm I}(N,M,e^{-4K})\right].
\end{equation}
Here
\begin{equation}
	\label{ppp-new}
	{\rm I}(N,M,z) :=\frac{4}{\pi}\int_0^{\pi/2}\cos(M x)\;
	T_{N}\left(\frac{\cos(x)}{\sqrt{1-z}}\right)dx,
\end{equation}
and
\begin{eqnarray}
	\label{babx-new}
	{\rm D}\left(N,M;z\right)=
	\frac{4}{\pi}\int_{0}^{\pi/2}\cos(M x)  \frac{\cos(x)}{\sqrt{1-z}}U_{N-1}\left(\frac{\cos(x)}{\sqrt{1-z}}\right)\,dx.
\end{eqnarray}
\eq{eq:ZD-in-CE-integral} can be obtained in an alternative, but more involved way, by plugging in \eq{sl2j}  the expressions for $ _2F_1\left(\frac{1}{2}(M-N+2),\frac{1}{2}(-M-N+2);2,z\right)$ and $ _2F_1\left(\frac{1}{2}(M-N+2),\frac{1}{2}(-M-N+2);1,z\right)$ derived in Ref. \cite{Dantchev2023b}. Indeed, these expressions can be obtained from the corresponding Eqs.(C.13), together with (C.15), and (C.37) (after correcting for the multiplication by a misprinted factor of  $N$) of Ref. \cite{Dantchev2023b}. Then these equations read
\begin{equation}
		\label{eq:I-expression}
		I(N,M,z)= N z (1-z)^{-N/2}\; _2F_1\left(\frac{1}{2}(M-N+2),\frac{1}{2}(-M-N+2);2,z\right),
\end{equation}
and 
\begin{eqnarray}
		\label{eq:D-expression}
		&&D(N,M,z)=(1-z)^{-N/2}z\; 
		\bigg\{N\; _2F_1\left(\frac{1}{2}(M-N+2),\frac{1}{2}(-M-N+2);2,z\right)+\nonumber\\
		&&2(z^{-1}-1)\;_2F_1\left(\frac{1}{2}(M-N+2),\frac{1}{2}(-M-N+2);1,z\right)\bigg\}. 
\end{eqnarray}

We comment that these expressions are interesting in their own right, being integral representations of Gaussian hypergeometric functions.

\section{Concluding remarks and discussion}
\label{sec:conclusion}

In this article we report on a study of two examples of fluctuation-induced forces: the Casimir force pertinent to $(T-h)$ grand canonical ensemble and the Helmholtz force in $T-M$ ensemble. We focused on the case of the one-dimensional Ising chain with Dirichlet boundary conditions. The Ising chain is one of the very few statistical-mechanical models which allow for an exact and explicit treatment of the behavior of those forces in the space of all physically realizable values of the thermodynamic variable. The main results are:

\begin{itemize}
	
	\item The existence of a nonzero external field $h\ne0$ is a necessary condition for the occurrence of a non-vanishing Casimir force in one dimensional Ising model with (free) Dirichlet boundary conditions
	(as opposed to the cases of PBC's and ABC's). This directly follows from \eq{eq:Casimir-nonzero-field} and is stated in 	\eq{eq:Casimir-at-zero-field}, in a full agreement with Ref. \cite{RZSA2010}, where only the case $h=0$ has been considered.
			 
	\item The Gibbs free energy and the behavior of the Casimir force within the grand canonical ensemble with Dirichlet BC's are derived in Sec. \ref{sec:free-energy-Casimir-forec}. The basic expression for the Casimir force is given in 	\eq{eq:Casimir-nonzero-field}, while its scaling behavior and the corresponding scaling function are given in \eq{eq:scaling-function-Casimir}. From these results it follows that the force is always \textit{attractive}.	As graphical images the behavior of the force as a function of the scaling variables $x_t$ and $x_h$ (see their definition in \eq{eq:scaling_variables_Ising}), as well as a function of $K$ and $h$, are presented in Fig. \ref{fig:3D-Casimir}. 
	
	\item Expressions for the Helmholtz free energy and the behavior of the Helmholtz force within the canonical ensemble with Dirichlet BC's are reported in Sec. \ref{sec:Helmholtz-force-and-energy}. Analytical results for the partition function are given in 	\eq{eq:DDbc-main-text} and \eq{sl2j}. The scaling behavior of the Helmholtz force is expressed in \eq{eq:Helmholtz-force} and \eq{eq:scaling-function-of-Helmholtz-force}. The behavior of the force as a function of $x_t$ and $m=M/N$ is shown in Fig.	\ref{fig:3D-Helmholtz}, while its behavior as a function fo $K$ and $m$ is depicted in Fig. \ref{fig:full-K-Helmholtz}. We observe that, contrary to the behavior of the Casimir force, the Helmholtz force changes sign as a function of $T$ and $m$ and can be \textit{both attractive and repulsive}. 
	\end{itemize}
	
	Technical details needed for the derivations and prove of our statements are organized in series of appendices, given at the end of the article. In some of them we made use of the properties of the Chebyshev polynomials and their relations to the Gauss hypergeometric functions that might be of interest of their own. As a byproduct we derived some relations between the  partition functions of the Ising chain---see \eq{SSR} for the relation between the partition functions under Dirichlet, periodic and antiperiodic boundary conditions in the GCE and the corresponding ones \eq{slb2} for the CE. 
	
	Finally, we stress that the fluctuation-induced forces play an important role in  nano-technologies - the general hope is that these forces can be used to manipulate nanoscale objects. The behavior of these forces in the different possible ensembles has not been subject of detailed studies up to now. The results reported here reinforce the conclusion that the standard rule for the Casimir force - that it is attractive, if the boundary conditions are similar, and repulsive otherwise, \textit{does not} hold for the Helmholtz force which can be both attractive and repulsive for all boundary conditions studies so far.  
	Our calculations reveal significant temperature dependence of the Helmholtz forces for a variety of boundary conditions, specifically  periodic, antiperiodic, or Dirichlet, leading to the possibility of a change in the nature of the force---attractive, or repulsive, contrary to  the corresponding Casimir force under the same boundary conditions,  which does not change sign  with temperature for these cases. 
	This points, at least in principle, to the possibility of a convenient dynamic control of the repulsive or attractive behavior of the critical Helmholtz force by varying an external thermodynamic parameter such as temperature.

Given the difference between
two ensembles for finite systems, it is reasonable to anticipate that the Casimir and Helmholtz force will have, in general, different
behavior for the same geometry and boundary conditions.

In light of  the new results for the behavior of the Helmholtz forces  of the one dimensional Ising model, an obvious question is: how generic it is? It seems  reasonable to anticipate similar behavior in the framework of  more complicated models, but such results are, as yet, lacking. 
Looking forward, it is reasonable to envision an interesting field for research that that is ripe for exploration, both analytically and numerically.

\appendix

\section{On the relation of $Z^{(D)}_{\rm GC}(N,K,h)$ to $Z^{\rm (per)}_{\rm GC}(N,K,h)$ and $Z^{\rm (anti)}_{\rm GC}(N,K,h)$}
\label{sec:GC-DD-relations}
thus

From ref. \cite{Dantchev2023b} 
for the partition functions with PBC's and ABS's we derived 
\begin{equation}
	\label{GCPBC}
	{Z}^{\rm (per)}_{\rm GC}(N,K,h)
	=2\left(\sqrt{2\sinh(2K)}\right)^N T_N\left(z(K,h)\right) 
\end{equation}	
and
\begin{eqnarray}
	\label{GCABC}
	{Z}^{\rm (anti)}_{\rm GC}(N,K,h)
	=2\left(\sqrt{2\sinh(2K)}\right)^N e^{-2K}z(K,h)U_{N-1}\left(z(K,h)\right),
\end{eqnarray}
respectively.
Collecting the partition functions  with different boundary conditions, i.e. Eqs.\eqref{GKD}, \eqref{GCPBC} and \eqref{GCABC}, we
obtain an expression for $Z^{(D)}_{GC}(N,K,h)$
through $Z^{\rm (per) }_{\rm GC}(N,K,h)$ and $Z^{\rm (anti)}_{\rm GC}(N,K,h)$:
\begin{eqnarray}
	\label{SSR}
	\cosh(h)\,Z^{(D)}_{\rm GC}(N,K,h)
	=\cosh^2(h) Z^{\rm (per)}_{\rm GC}(N-1,K,h)+
	\,[e^{2K}\sinh^2(h)+1]\,Z^{\rm (anti)}_{\rm GC} (N-1,K,h).
\end{eqnarray}

If we set $h=0$ in Eq.\eqref{SSR} we obtain a remarkably simple relation
\begin{eqnarray}
	\label{SSRaa}
	Z^{(D)}_{{\rm GC}}(N,K,0)
	= Z^{\rm (per)}_{\rm GC}(N-1,K,0)+
	\,Z^{\rm (anti)}_{\rm GC} (N-1,K,0).
\end{eqnarray}

\section{Proof of the  inequality $0<A(K,h)<\cosh(h)$ }	
\label{Ineq}

$A(K,h)$ is defined in	\eq{eq:A-definition}. 
We have to prove that 
\begin{equation}
	0<\frac{\sinh^2(h)+e^{-2K}}{\sqrt{\sinh^2(h)+e^{-4K}}}<\cosh(h), \quad \mbox{when} \quad K>0, \quad \mbox{and} \quad h \ne 0. 
\end{equation}
Proof:

The lhs is evident. The rhs may be rewritten in the form:
\begin{equation}
e^{-K}\frac{x+1}{\sqrt{x+e^{-2K}}}<
	\sqrt{1+e^{-2K}x},\quad \mbox{where}\quad x:=\sinh^2(h)/e^{-2K}\geq 0.
\end{equation}
Then
\begin{equation}
	\frac{(x+1)^2}{1+x \, e^{2K}}<
	1+e^{-2K} x.
\end{equation}		
Finally, the above is valid since
\begin{equation}
	1<
	\cosh(2K).\qquad\blacksquare
\end{equation}

\section{Rationalized presentation of the canonical partition function in terms of Gauss hypergeometric functions}
\label{rpf}

Here we demonstrate how one can derive \eq{sl2j} in terms of the Gauss functions.
For simplicity, let us introduce the notations 
\begin{eqnarray}
	\label{not}
	\alpha:=\frac{1}{2}(M-N+2),\quad \alpha^+:=\frac{1}{2}(M-N+4),\nonumber\\
	\quad \beta:=\frac{1}{2}(-M-N+2),\quad
	\beta^+:=\frac{1}{2}(-M-N+4).
\end{eqnarray}
in the rhs of \eq{eq:DDbc-main-text}. The result is
\begin{align}
	\label{sl}
	& Z^{(D)}_C(N,K,M)= e^{K(N-1)}  \nonumber\\
	& \times \bigg\{2e^{-2K}\,_2F_1(\beta,\alpha;1;e^{-4K})
	-\,e^{-4K}\bigg[\alpha\,_2F_1(\alpha^+,\beta;2;e^{-4K})
	+\beta\,_2F_1(\alpha,\beta^+;2;e^{-4K})\bigg]\bigg\},
\end{align}
where the property 
\begin{equation}
	\label{sym}
	_2F_1(\alpha,\beta;\gamma,z)=\,_2F_1(\beta,\alpha;\gamma,z) 
\end{equation}
has been used.

The relations between hypergeometric function $\,_2F_1(\alpha,\beta,;\gamma;z) $ and any two  hypergeometric functions with the same argument "z" and with parameters  $\alpha, \beta $ and $\gamma$  changed by $\pm 1$ are termed {\it contiguous} relations.
In order to transform the functions in the second line in Eq.\eqref{sl} we will use the contiguous  relations \cite[see there Eq. (1.5) with $c=2$ ]{Rakha2011}:
\begin{equation}
	\alpha\,_2F_1(\alpha^+,\beta;2;z)=(\alpha-1)\,_2F_1(\alpha,\beta;2;z) + \,_2F_1(\alpha,\beta;1;z)
\end{equation}
and
\begin{equation}
	\beta\,_2F_1(\alpha,\beta^+;2;z)=(\beta-1)\,_2F_1(\alpha,\beta;2;z) + \,_2F_1(\alpha,\beta;1;z).
\end{equation}
Once we add together the both equations and set the result in Eq.\eqref{sl} we get 
\begin{align}
	\label{sl32}
	& Z^{(D)}_C(N,K,M)= \\
	& 2e^{K(N-3)}\,_2F_1(\beta,\alpha;1;e^{-4K})
	-e^{K(N-5)}\bigg[(\alpha+\beta-2)\,_2F_1(\alpha,\beta;2;e^{-4K}) + 2 \,_2F_1(\alpha,\beta;1;e^{-4K}) \bigg], \nonumber
\end{align} 
or finally
\begin{eqnarray}
	\label{sl23}
	Z^{(D)}_C(N,K,M)= 4e^{K(N-4)}\sinh(K)\,_2F_1(\alpha,\beta;1;e^{-4K})
	+N e^{K(N-5)}\,_2F_1(\alpha,\beta;2;e^{-4K}). 
\end{eqnarray} 

Returning to the original notation  in the Gauss hypergeometric functions, see \eq{not}, we obtain \eq{sl2j} reported in the main text.

\section{On the relation of $Z^{(D)}_{\rm C}(N,K,M)$ to $Z^{\rm (per)}_{\rm C}(N,K,M)$ and  $Z^{\rm (anti)}_{\rm C}(N,K,M)$}
 
Using Eqs. (2.19), (2.20) and (2.21) from \cite{Dantchev2023b} for the partition function with PBC's one has:
  \begin{equation}
 	\label{eq:Z-fixed-m-via-angle-pi-over-two2}
 	Z^{(\rm per)}_C(N,K,M) =\bigg(\sqrt{2 \sinh (2 K)}\,\bigg)^N	 {\rm I}(N,M;e^{-4K})
 \end{equation}
 while from Eqs. (3.20) (3.21) and (3.22) for  ABC's
 \begin{eqnarray}
 	\label{eq:transfer-matrix-antiperiodic-final}
 	Z^{(\rm anti)}_C(N,K,M) = \bigg(\sqrt{2\sinh(2K)}
 	\,\bigg)^Ne^{-2K}  {\rm D}(N,M;e^{-4K}).
\end{eqnarray}
 Expressing from \eq{eq:Z-fixed-m-via-angle-pi-over-two2} and \label{eq:transfer-matrix-antiperiodic-final} ${\rm I}(N,M;e^{-4K})$ and ${\rm D}(N,M;e^{-4K})$,  from \eq{eq:ZD-in-CE-integral} one concludes that
 \begin{equation}
 	\label{slb2A}
 	Z^{(D)}_C(N,K,M)=\frac{1}{2 \cosh(K)}\left[
 	Z^{(\rm per)}_C(N,K,M) +Z^{(\rm anti)}_C(N,K,M)\right]
 \end{equation}

 \section{Proof of the equality of Eqs. \eqref{Zd} and \eqref{eq:ZD-in-CE-integral} }
 \label{AF}
 Here we prove that rhs of Eq.\eqref{Zd} equals the rhs of Eq.\eqref{eq:ZD-in-CE-integral}. For this purpose we need the following recurrence relations among the Chebyshev polynomials:
 \begin{equation}
 	\label{Ch1}
 	T_{N-1}(x)= U_{N-1}(x)-x U_{N-2}(x)
 \end{equation} 
 and
 \begin{equation}
 	\label{CH2}
 	U_{N-2}(x)=x U_{N-1}(x)-T_N(x),
 \end{equation}
which can be easily obtained from \cite[8.941.1-8.941.3]{GR}. With the help of \eq{Ch1}) the  integrand  in the first term in the rhs of \eq{Zd} may be rewritten  in the form 
 \begin{eqnarray}
 	\label{cuk}
 	&&\cos(\phi)\cos( M \phi) T_{N-1}\bigg(\tilde{z}(K,\phi\bigg)=
 	\cos(M\phi)\sqrt{1-e^{-4K}}\bigg(\tilde{z}(K,\phi)\bigg)T_{N-1}\bigg(\tilde{z}(K,\phi)\bigg)=\nonumber\\
 	&&\cos(M\phi)\sqrt{1-e^{-4K}}\bigg\{\tilde{z}(K,\phi)U_{N-1}(\tilde{z}(K,\phi))-\tilde{z}^2(K,\phi)U_{N-2}(\tilde{z}(K,\phi)) \bigg\}.
 \end{eqnarray}
 Using Eq. \eqref{CH2}, the  integrand  in the second term in the rhs of \eq{Zd} may be cast in the form 
 \begin{eqnarray}
 \label{poy}
 \bigg[\frac{e^{-2K}-1}{\sqrt{1-e^{-4K}} }
 \bigg]
 \cos ( M \phi)U_{N-2}
 \bigg(\tilde{z}(K,\phi)\bigg)+
 \cos ( M \phi)\sqrt{1-e^{-4K}}
 	\tilde{z}^2(K,\phi)U_{N-2}
 	\bigg(\tilde{z}(K,\phi)\bigg).
 	\end{eqnarray}
 The last terms in Eqs.	\eqref{cuk} and \eqref{poy} cancel  each other. With the help of Eq.\eqref{CH2} for the remaining term in Eq.\eqref{poy} we obtain 
 	\begin{eqnarray}
 \label{bahe}
 \bigg[\frac{e^{-2K}-1}{\sqrt{1-e^{-4K}} }
 \bigg]
 \cos ( M \phi)U_{N-2}
 \bigg(\tilde{z}(K,\phi)\bigg)=
 \bigg[\frac{e^{-2K}-1}{\sqrt{1-e^{-4K}} }
 \bigg]
 \cos(M\phi)\bigg\{\tilde{z}(K,\phi)U_{N-1}(\tilde{z}(K,\phi))-T_N(\tilde{z}(K,\phi))\bigg\}.\nonumber\\
 \end{eqnarray}
 Combining the first term  in \eqref{cuk} with the first term in \eqref{bahe} we get
 \begin{eqnarray}
 &&	\cos(M\phi)\sqrt{1-e^{-4K}}\bigg\{\tilde{z}(K,\phi)U_{N-1}(\tilde{z}(K,\phi))\bigg\} +\bigg[\frac{e^{-2K}-1}{\sqrt{1-e^{-4K}} }
 \bigg]
 \cos ( M \phi)\bigg\{\tilde{z}(K,\phi)U_{N-1}(\tilde{z}(K,\phi))\bigg\}=\nonumber\\
 &&	\cos(M\phi)
 \frac{1-e^{-2K}}{\sqrt{1-e^{-4K}}}
 e^{-2K}
 \bigg\{\tilde{z}(K,\phi)U_{N-1}(\tilde{z}(K,\phi))\bigg\}
 \end{eqnarray}
 Plugging the above expression in  Eq.\eqref {Zd}  and taking into account the definition Eq.\eqref{babx-new} we obtain the first term in Eq.\eqref{eq:ZD-in-CE-integral}. The second term in Eq.\eqref{bahe} is
 \begin{equation}
 \bigg[\frac{e^{-2K}-1}{\sqrt{1-e^{-4K}} }
 \bigg]
 \cos(M\phi)T_N(\tilde{z}(K,\phi)),
 \end{equation}
 which in conjunction  with the definition Eq.\eqref{ppp-new} provides the second term in Eq.\eqref{eq:ZD-in-CE-integral}.
 This finishes the proof of  the result stated. $\qquad \blacksquare$

  \section*{Acknowledgments}
  The partial financial support via Grant No KP-06-H72/5 of Bulgarian NSF is gratefully acknowledged.

 \end{document}